\RequirePackage{fix-cm}
\documentclass[twocolumn, epjc3, final]{svjour3}

\usepackage{atlasphysics}    
\usepackage{mydefs}             
\usepackage[hyperindex,breaklinks]{hyperref} 
\usepackage{graphicx}
\usepackage{mathptmx} 
\usepackage{subfigure}
\usepackage[figuresleft]{rotating}
\usepackage{epstopdf}
\usepackage[switch]{lineno}
\usepackage{float}
\usepackage[parsep]{collref}
\usepackage{hyperref} 
\usepackage{cite} 
\usepackage{xspace}
\usepackage{placeins}
\usepackage{dblfloatfix} 
\smartqed       

\makeatletter
\g@addto@macro\bfseries{\boldmath}
\makeatother

\journalname{Eur.\ Phys.\ J.\ C}

\usepackage{preprintcover}  % load package
\PreprintCoverPaperTitle{Search for heavy neutrinos and right-handed $W$ bosons
	 in events with two leptons and jets in $pp$ collisions at $\sqrt{s}$ = 7~\TeV\
 	 with the ATLAS detector}  % paper title
\PreprintIdNumber{CERN-PH-EP-2012-022}  % CERN preprint number
\PreprintCoverAbstract{This letter reports on a search for 
hypothetical heavy neutrinos, $N$, and right-handed gauge bosons, \WR, 
in events with with high transverse momentum objects 
which include two reconstructed leptons and at least one hadronic jet.
The results were obtained from data corresponding to an 
integrated luminosity of $2.1~\ifb$
collected in proton-proton collisions at \rts=7~\TeV\  
with the ATLAS detector at the CERN Large Hadron Collider. 
No excess above the Standard Model background expectation is observed. 
Excluded mass regions 
for Majorana and Dirac neutrinos are presented using two approaches 
for interactions that violate lepton and lepton-flav\-our numbers.
One approach uses an effective operator framework, 
the other approach is guided by the Left-Right Symmetric Model. 
The results described in this letter represent the most stringent limits to date 
on the masses of heavy neutrinos and \WR\ bosons obtained in direct searches.}  % paper abstract
\PreprintJournalName{Eur. Phys. J. C}  % journal name

\begin{document}

%%%%%%%%%%%%%%%%%%%%%%%%
% Title
%%%%%%%%%%%%%%%%%%%%%%%%
\title{Search for heavy neutrinos and right-handed $W$ bosons
	 in events with two leptons and jets in $pp$ collisions at $\sqrt{s}$ = 7~\TeV\
 	 with the ATLAS detector }
\author{The ATLAS Collaboration}
\authorrunning{The ATLAS Collaboration} 
\institute{CERN, 1211 Geneva 23, Switzerland \\ \email{atlas.publications@cern.ch}}
%\date{Received: \today / Accepted: date Done by editor}   % The correct dates will be entered by the editor
\date{\today}   
\titlerunning{Search for heavy neutrinos and \WR}

%%%%%%%%%%%%%%%%%%%%%%%%%
% Abstract
%%%%%%%%%%%%%%%%%%%%%%%%%
\maketitle
\abstract{
%This letter reports on a search for 
%hypothetical heavy neutrinos, $N$, and right-handed gauge bosons, \WR, 
%in events with two high transverse momentum leptons 
%and at least one high transverse momentum hadronic jet. 
This letter reports on a search for 
hypothetical heavy neutrinos, $N$, and right-handed gauge bosons, \WR, 
in events with high transverse momentum objects 
which include two reconstructed leptons and at least one hadronic jet.
The results were obtained from data corresponding to an 
integrated luminosity of $2.1~\ifb$
collected in proton-proton collisions at \rts=7~\TeV\  
with the ATLAS detector at the CERN Large Hadron Collider. 
No excess above the Standard Model background expectation is observed. 
Excluded mass regions 
for Majorana and Dirac neutrinos are presented using two approaches 
for interactions that violate lepton and lepton-flav\-our numbers.
One approach uses an effective operator framework, 
the other approach is guided by the Left-Right Symmetric Model. 
The results described in this letter represent the most stringent limits to date 
on the masses of heavy neutrinos and \WR\ bosons obtained in direct searches.
%Non-standard-model neutrinos, right-handed neutrinos, etc. ,Experimental test, limits on production of particles
}

%%%%%%%%%%%%%%%%%%%%%%%%%
% Paper sections
%%%%%%%%%%%%%%%%%%%%%%%%%
\section{Introduction}
\label{Sec:Intro}

The discovery of neutrino oscillations~\cite{Review, KamLand}
unambiguously establishes that neutrinos have non-zero mass 
and provides clear evidence for physics beyond the Standard Model (SM).
One possible explanation for the mass of light neutrinos 
is provided by theoretical models based on a Grand Unified Theory (GUT). 
Such models often introduce one or more additional neutrino fields,
which manifest themselves as new heavy particles 
that could be directly observable at the Large Hadron Collider (LHC). 
In the framework of GUT models, the mass of the light neutrinos 
could be explained via the see-saw mechanism~\cite{Mohapatra:1979ia,Mohapatra:1980yp,Davidson:1987tr,AAtre:2009}. 
This predicts $m_\nu\approx m_D^2/m_N$, 
where, for each generation, 
$m_\nu$ is the mass of a known light neutrino,
$m_D$ is the Dirac mass for charged fermions of the same generation, 
and $m_N$ is the mass of a new heavy neutrino, $N$. 
If the see-saw mechanism were to explain the mas\-ses of the known neutrinos, 
both the light and the heavy neutrinos would have to be Majorana particles.  
This would violate lepton number conservation, 
and yield a striking signature of two leptons with the same charge at the LHC~\cite{Keung:1983uu}.

This letter reports on a search for new heavy neutrinos of either Majorana or Dirac-type, 
with data corresponding to an integrated luminosity of $2.1~\ifb$ recorded with the ATLAS detector at the LHC. 
Two approaches are employed. 
The first approach aims at exploring possible sources of new physics predicting
heavy neutrinos 
using a Lagrangian of effective operators (referred to as HNEO hereafter)~\cite{theoryInclusiveHn}. 
The theory is built on effective four-fermion operators ($q\bar q'\rightarrow N \ell $) 
with the N decaying promptly via a three-body decay, $N\rightarrow \ell jj$.
The second approach is based on the concept of Left-Right Symmetry~\cite{Pati:1974yy,Mohapatra:1974gc,Senjanovic:1975rk}
which extends the electroweak part of the SM by a new gauge group. 
Its force particles (\WR\ and $Z^\prime$ bosons) could be produced at LHC energies. 
A particular implementation of left-right symmetry breaking~\cite{Mohapatra:1980qe}, the 
Left-Right Symmetric Model (LRSM) with doubly-charged Higgs bosons~\cite{Ferrari:2000sp,model_LRSM}  
is used in the present analysis. 
According to this model, the heavy neutrinos would be produced in the decays of a \WR\ boson via 
$q\bar q' \rightarrow \WR \rightarrow \ell N$, with $N$ decaying subsequently via 
$N \rightarrow \ell W^{*}_{\mathrm{R}} \rightarrow \ell j j$. 
Thus, the final state signature for both models consists of two leptons and two jets with high transverse momenta ($p_{\rm T}$).
Only electrons and muons are considered in this analysis.

The $N$ invariant mass can be fully reconstructed from the decay products in both approaches.
Given the s-channel production in the LRSM, the \WR\ mass, $m_{\WR}$, can also be reconstructed in this model.
The reconstructed \WR\ boson and $N$ masses are used to perform 
the search in the context of the HNEO and LRSM models, respectively.
Like the SM neutrinos, heavy neutrinos can mix  if their masses are different. 
Both the scenarios of no-mixing~\cite{Tello:2010am} and maximal mixing~\cite{Melfo:2011nx} 
between two generations of lepton flavours (electron and muon) are investigated assuming that 
the mass difference between the heavy neutrinos is much smaller than 
the experimental resolution of their reconstructed invariant mass.  
In the case of maximal mixing, a mass difference of $2~\GeV$ is assumed.
If the heavy neutrinos are of Majorana type, they would contribute to both the same-sign (SS)
and opposite-sign (OS) channels, while heavy Dirac neutrinos would contribute solely to the OS channel.

Heavy neutrinos were previously searched for at LEP and excluded for masses up to 
$\approx 100\GeV$~\cite{LEP2ALEPH,LEP2DELPHI,LEP2L3,LEP2OPAL}. 
The most stringent direct limits on \WR\ bosons~\cite{WRD0,WRCDF}
come from the Tevatron, where $\WR \rightarrow tb$ decays were searched for. 
Assuming a branching ratio of 100\%, \WR\ boson masses below $825\GeV$ are excluded at 95\% confidence level (C.L.). 
Recently, the ATLAS collaboration published an inclusive search for new 
physics in the same-sign dilepton signature for an integrated luminosity of 34~\ipb~\cite{exoSSIncl}.
The 95\% C.L. limits presented exclude \WR\ masses up to about 1~\TeV\ for the LRSM model and
Majorana neutrino masses around \\
460~\GeV\ for the HNEO model.

\section{The ATLAS detector}
\label{Sec:Det}

The ATLAS detector~\cite{ATLASExp} is a multipurpose particle physics apparatus with a forward-backward 
symmetric cylindrical geometry and nearly 4$\pi$ coverage in solid angle\footnote{ATLAS uses a right-handed 
coordinate system with its origin at the nominal interaction point in the centre of the
detector and the $z$-axis coinciding with the axis of the beam pipe. 
The $x$-axis points from the interaction point 
to the centre of the LHC ring, and the $y$-axis points upward. 
Cylindrical coordinates ($r$,$\phi$) are used in the transverse plane, $\phi$ being the azimuthal angle around
the beam pipe. The pseudorapidity is defined in terms of the polar angle $\theta$ as $\eta = {\rm -ln} \tan(\theta/2)$.}. 
The inner tracking detector (ID) covers the pseudorapidity range $|\eta|<2.5$ and consists of: a silicon pixel detector, 
providing typically three measurements per track; 
a silicon microstrip detector (SCT), which provides typically four to five measurements; 
and, for  $|\eta|<2.0$, a transition radiation tracker (TRT), giving typically 30 straw-tube measurements per track.
The ID is surrounded by a thin superconducting solenoid providing a 2~T magnetic field.
A high-granularity liquid-argon (LAr) sampling electromagnetic calorimeter covers the region $|\eta|<3.2$.
An iron-scinti\-llator tile calorimeter provides hadronic coverage in the central rapidity range of $|\eta|<1.7$. 
The end-cap and forward regions, spanning $1.5<|\eta|<4.9$, are instrumented with LAr calorimeters for both 
electromagnetic and hadronic measurements. The muon spectrometer (MS) surrounds the calorimeters and 
consists of a system of air-core superconducting toroid coils,
precision tracking chambers up to $|\eta|<2.7$, and detectors for triggering in the region of $|\eta|<2.4$.

\section{Trigger and Data}
\label{Sec:TrigData}

The data used in this analysis were recorded between March and August 2011 at a centre-of-mass energy of 7 TeV. 
The application of beam, detector, and data quality requirements results in a total integrated luminosity of 2.1~\ifb
with an estimated uncertainty of $\pm 3.7\%$~\cite{lumiPaper, lumiConf}.
The data were recor\-ded with single lepton ($e$ or $\mu$) triggers~\cite{ATLAS_Trigger}.
At the last stage of the trigger decision, the electron trigger selects candidate
electrons with transverse energy $E_{\rm T}>20\GeV$, satisfying shower-shape requirements
and matching an ID track. 
For the last part of the dataset, corresponding to an integrated luminosity of 0.5~\ifb, 
the threshold was raised to 22~\GeV.
The muon trigger selects candidate muons with $p_{\rm T}>18\GeV$ and $|\eta|<2.4$.
These triggers reach full efficiency for electrons with $p_{\rm T}>25~\GeV$ and muons with $p_{\rm T} >20~\GeV$.
The typical trigger efficiencies measured from data for leptons selected for offline analysis are $99\pm1 \%$ 
for electrons, and 74\% and 91\% for muons in the barrel ($|\eta|<1.05$) 
and end-cap ($1.05<|\eta|<2.4$) regions, respectively, 
with an uncertainty of about $\pm 1\%$.

\section{Monte Carlo Simulation}
\label{Sec:DataMc}

Fully simulated Monte Carlo (MC) event samples are used to develop and validate the analysis procedure, 
estimate the detector acceptance and reconstruction efficiency, and aid in the background determination.
The simulation of background processes is described in detail in Ref.~\cite{TopXs}.
For the major backgrounds, $Z/\gamma^*$+jets production and top quark pair production, \alpgen~\cite{ALPGEN} 
and \mcatnlo~\cite{mcatnlo} are used, respectively. 
The leading-order parametrisation~\cteqVER{6L1}~\cite{Cteq6} of the parton density functions (PDF) 
is used for the \alpgen\ simulation, while the next-to-leading order parametrisation\\
 \cteqVER{6.6}~\cite{Cteq6} is used for the \mcatnlo\ simulation.
Fragmentation and hadronization are performed in both cases with 
\herwig~\cite{Herwig}, using \jimmy~\cite{Jimmy} for the underlying event modelling.
Diboson ($WW$, $WZ$, and $ZZ$) event samples are generated using \herwig, 
while \madgraph~\cite{MadGraph} interfaced to \pythia~\cite{Pythia} is used for $W\gamma$ and $Z\gamma$ production. 
Single top-quark production is generated with \mcatnlo.
The production of $W^+W^+$ arising from a t-channel 
gluon exchange, resulting in two jets in the final state and two same-sign $W$ bosons, are generated with
\madgraph\ interfaced to \pythia. 
The associated production of a vector boson with a \ttbar\ pair ($\ttbar W$, $\ttbar Z$, $\ttbar \gamma$) 
is simulated with \madgraph\ interfaced with \pythia.

The HNEO signal MC samples are generated using\\
\calchep~\cite{CalcHEPGen} and the leading-order 
PDF \cteqVER{6L}~\cite{Cteq6}, and hadronization simulated with \pythia.
All lepton combinations of $e,\mu$ or $\tau$ leading to lepton number violating (LNV) signatures, which produce
SS or OS dilepton events, are included.
The model is implemented via a Lagrangian of effective operators defined as
\begin{equation}
  {\mathcal L} = \sum_{n=5}^{\infty} \frac{1}{\Lambda^{n-4}} \cdot 
  \sum_{i} \alpha_{i} {{{\mathcal O}_{i}}^{(n)}},
\end{equation}
\noindent where $n$ is the operator dimension, $\Lambda$ is the scale of LNV interactions, 
$\alpha_i$ are the coupling constants between the neutrino $N$ and the leptons, and ${\mathcal O_i}$ are the 
effective operators~\cite{theoryInclusiveHn}.
The signal samples are produced for four effective operator hypotheses: 
the four-fermion vector operator, $\mathcal{O}_V$, and
four-fermion scalar operators, $\mathcal{O}_{s1}$, $\mathcal{O}_{s2}$, 
and $\mathcal{O}_{s3}$. 
The tree-level-generated dimension-6 operator $\mathcal{O}_V$ corresponds to $duNe$, while 
$\mathcal{O}_{s1}$ and $\mathcal{O}_{s2}$ correspond to $QuNL$ and $LNQd$ respectively, and
$\mathcal{O}_{s3}$ corresponds to $QNLd$ ($e$, $u$, $d$ and $L$, $Q$ denote 
the right-handed SU(2) singlets and left-handed SU(2) doublets, respectively).
The production via the effective operators $\mathcal{O}_{s1}$ and $\mathcal{O}_{s2}$ 
have the same cross section and lead to identical event kinematics, 
which makes them indistinguishable. 
The production cross sections for the Majorana and Dirac neutrinos in 
the framework of the effective Lagrangian are related to the energy scale of
new physics and the coupling constant $\sigma \approx \alpha^2/\Lambda^4$, 
such that the coupling can be varied to scan for new physics at different $\Lambda$ scales.

The LRSM signal MC samples 
are generated using an implementation of this model~\cite{model_LRSM} in \pythia, 
with modified leading-order parton distribution functions \mrstlo\\~\cite{PDF}. 
The coupling constants for the \WR\ and left-handed $W$ boson are assumed to be the same, 
including the CKM matrix for \WR\ boson couplings to right-handed chiral quark components. 
It is assumed that there is no mixing between the \WR\ boson and the SM $W$ boson. 
The LRSM signal MC samples are generated constraining the decays to $e$ or $\mu$ and 
with $m_N<m_{\WR}$. The branching fractions used are the ones predicted by \pythia.
When the mass difference between the \WR\ and the $N$ is large, the leptonic branching fractions are $\approx 8\%$, 
and they decrease with decreasing mass difference.

Both Majorana and Dirac type heavy neutrinos are considered, 
assuming that the total production cross section is the same for both cases. 
The leading-order theoretical cross sections are used. 

All signal and background samples are generated using the ATLAS underlying event tunes~\cite{pythiaTune, jimmyTune} and
processed through the ATLAS detector simulation~\cite{ATLASSim} based on \geant~\cite{geant4}.
The MC samples are produced including the simulation of multiple interactions per LHC bunch crossing (pile-up). 
Varying pile-up conditions and their dependence on the instantaneous luminosity of the LHC are taken
into account by reweighting MC events to match the pile-up conditions measured in data.

\section{Object reconstruction and event selection}
\label{Sec:EvtSel}

The criteria for electron and muon identification closely follow those described in Ref.~\cite{WZXs}. 
Electrons are required to pass the ``medium'' selection criteria,
with $p_{\rm T}>25\GeV$ and $|\eta|<2.47$, excluding the electromagnetic
calorimeter transition region, $1.37<|\eta|<1.52$~\cite{ElePerf}. To improve the background
rejection for $|\eta|>2.0$, more stringent requirements are placed
on the track-cluster matching in $\eta$ and shower shape.
Electron tracks that pass through an active region of the innermost
pixel detector are required to have a measurement in that layer in order to suppress
electrons from photon conversions. 
Additionally, an electron whose track matches the ID track of a muon candidate is rejected.

Muons are required to be identified in both the ID and the MS systems.
The ID track is required to have at least one pixel hit, more than five SCT hits, 
and a number of TRT hits that varies with $\eta$.
Muon tracks that pass
through an active region of the innermost pixel detector are required to have a measurement in that layer.
The curvatures, as measured by the ID and MS systems, must have 
the same sign.  
Only muons with $p_{\rm T}>25\GeV$ and $|\eta|<2.4$ are considered.
Selection criteria on the displacement of the muon relative to the primary vertex, 
selected as the one with the highest $\sum p_{\rm T}^2$ of associated tracks, are required.
The longitudinal ($z_0$) and transverse ($d_0$) impact parameters must satisfy 
 $|z_0|<5$ mm,  $|d_0|<0.2$ mm, and $|d_0/\sigma_{d_0}|<5$, where $\sigma_{d_0}$ is the uncertainty on $d_0$. 
These cuts reduce the cosmic ray muon background to a negligible 
level and also reduce the background from non-prompt muons\footnote{Leptons from $W$, $Z$ and $\tau$ 
decays are classified as prompt leptons, while leptons any hadron decays are classified as non-prompt leptons.}.
 
To reduce the background due to leptons from decays of hadrons (including heavy-flavour hadrons) produced
in jets, requirements on the isolation of leptons are imposed. 
To evaluate the isolation energy for electrons, the transverse energies deposited in the calorimeter
towers in a cone in $\eta$-$\phi$ space of radius
 $\Delta R=\sqrt{(\Delta \phi)^2 + (\Delta \eta)^2}=0.2$ around the electron 
direction are summed and corrected for energy deposition from pile-up events. 
In addition,  the transverse energy of the electron, $E_{\rm T}$, 
corrected for energy leakage into the neighboring towers, is subtracted. 
The isolation transverse energy is required to be less than 15\% of the electron $E_{\rm T}$.
An equivalent quantity, $E_{\rm T}^{\Delta R=0.3}$, is calculated for the muon using a cone size of $\Delta R=0.3$.
If there is no jet with $\pT>20~\GeV$ within $\Delta R<0.4$ of the muon, $E_{\rm T}^{\Delta R=0.3}$ is required to be less than 15\% 
of the muon \pT. Additionally, muons with $\pT<80~\GeV$ should have no other track with $\pT>1~\GeV$ originating from the primary vertex within a cone of $\Delta R=0.3$ around the muon.
Otherwise, if the muon has a jet nearby, it must satisfy $\pT>80~\GeV$ and 
$(E_{\rm T}^{\Delta R=0.3}/\pT\\-3)/\pT>-0.02~\GeV^{-1}$.
These isolation requirements are powerful in rejecting background muons 
and highly efficient for selecting signal muons 
produced in the decays of heavy neutrinos 
and reconstructed near the signal jets 
in cases where the heavy neutrino is boosted. 

Jets are reconstructed using the anti-$k_{\rm t}$ jet clustering algorithm~\cite{antiKt, fastjet}
with a radius parameter $R=0.4$. The inputs to this algorithm are clusters of calorimeter
cells seeded by cells with energies significantly above the measured noise. 
The energies and momenta of jets are evaluated by performing 
a four-vector sum over these clusters, treating each cluster 
as an ($E$, $\vec{p}$) four-vector with zero mass. 
Jets are corrected for calorimeter non-compensation, upstream material and other effects using $p_{\rm T}$ and 
$\eta$-dependent calibration factors~\cite{jetVal} obtained from MC simulation~\cite{NI},
and validated with test-beam and collision-data studies.
Only jets with $p_{\rm T}>20\GeV$ and $|\eta|<2.8$ are considered. 
To avoid double counting, the closest jet within $\Delta R<0.5$ of an electron candidate is discarded.
The selected jets must pass quality requirements
based on their shower shape, and their calori\-meter signal timing must be consistent 
with the timing of the beam crossings~\cite{JES}. 
Events with any jet that fails the jet quality criteria are rejected.
To suppress jets unrelated to the hard scattering of interest, 
at least 75\% of the summed \pT\ of all reconstructed tracks 
associated with a jet with $|\eta|<2.8$ must come from tracks originating 
from the selected primary vertex. 
During a part of the data-taking period, corresponding to an integrated luminosity of 0.9~\ifb, 
an electronic failure in a small $\eta-\phi$ region of the LAr EM calorimeter created
a dead region.
For this integrated luminosity, events in data and MC containing either an identified electron or
a jet, with $\pT >40~\GeV$, satisfying
$-0.1<\eta<1.5$ and $-0.9<\phi<-0.5$ are rejected, leading to a loss of signal efficiency of about 10\% for this portion of the data.

Events are preselected by requiring exactly two identified leptons with $p_{\rm T}>25\GeV$ originating from the primary vertex and  
at least one jet with $p_{\rm T}>20\GeV$. 
At least one of the lepton candidates must match a triggered lepton at the last stage of the trigger selection.
To reduce the number of background events from Drell-Yan production
and misidentified leptons, 
the dilepton invariant mass, \mll, is required to be greater than 110~\GeV. 
The signal region is then subdivided into SS and OS dilepton events.
In the OS dilepton channels, further background reduction is achieved by requiring that the 
scalar sum of the transverse energies of the two leptons and the leading two jets with $p_{\rm T}>20\GeV$, denoted by $S_{\rm T}$,
is greater than 400~\GeV. 
This event selection is referred to hereafter as the baseline selection.
As mentioned previously, the mass of the $N$ can be reconstructed from its decay products of one lepton and two jets.
In the case where the N is boosted, the hadronic decay products can be reconstructed as one jet due to their proximity to each other. 
For scenarios with a large mass splitting between the \WR\ and $N$, up to half of the signal events have only one jet.
The \WR\ boson invariant mass is reconstructed from the leptons and the two highest $p_{\rm T}$ jets 
in events with at least two jets, or a single jet in events with only one jet. 
Anti-$k_{\rm t}$ jets are massive, and therefore, the jet four-momenta are used in calculating the invariant mass.
For the LRSM, the \WR\ boson invariant mass, \mlljj, is required to be greater than 400~\GeV\ for both SS and OS final states.

\section{Background estimation}
\label{Sec:Bkg}

Several processes have the potential to contaminate the signal regions.
The main background to the SS dilepton final state, which is referred to as ``fake lepton'' background,
arises from SM $W$+jets,  $t\bar t$, and multi-jet production 
where one or more jets are misidentified as prompt isolated leptons.
This background is measured using a data-driven technique rather than using less accurate estimates from MC simulation.
The other significant background arises from
charge misidentification of a reconstructed electron as a result of hard brem\-sstrahlung followed by asymmetric conversion \\
($\it e^\pm_{\rm hard}  \rightarrow  e^\pm_{\rm soft} \gamma_{\rm hard}^{} \rightarrow e^\pm_{\rm soft} e^\pm_{\rm soft} e^\mp_{\rm hard}$).
This background is estimated with a combination of MC and data-driven techniques.
Small contributions from diboson and single top-quark events are also accounted for using MC.

For the $e^{\pm}e^{\mp}$ and $\mu^{\pm}\mu^{\mp}$ final states, the dominant backgrounds are $Z/\gamma^*$+jets 
and \ttbar\ events, with about equal contributions after all selection criteria are applied.
The $e^{\pm}\mu^{\mp}$ final state is dominated by \ttbar\ production. 
The backgrounds from \ttbar, single top-quark, and diboson production are estimated from MC simulation, 
while the estimation of the $Z/\gamma^*$+jets  background is extracted from a normalization to the data. 
The fake lepton background is estimated from data, using the same method as for the SS final states.

A data-driven approach, similar to the one described in Refs.~\cite{exoSSIncl, TopXs}, is used to estimate the fake lepton background.
The method uses ``loose'' leptons in addition to the candidate leptons. 
Loose muons are defined using the same identification criteria as the candidate muons, 
except for the isolation requirements, which are not applied. For the electrons, looser requirements on the shower 
shape variables, track-cluster matching, and track quality are used, and the isolation requirement is not applied.
The method uses the fractions of these loose fake leptons, $R_{\rm fake}$, and loose prompt leptons, $R_{\rm prompt}$,
which also pass the candidate lepton requirements.  
A $4\times 4$ matrix is then employed on the ``loose-loose'' and ``loose-tight'' dilepton sample to predict the total 
fake lepton background contributing to the SS and OS dilepton final states.
The $R_{\rm fake}$ fractions are measured using fake lepton enriched control regions containing a single loose lepton.
Additional criteria are imposed to reduce the true lepton contamination 
from electroweak processes to a negligible level. 
For events in the control regions, the transverse mass,
$m_{\rm T}=\sqrt{2\cdot p_{\rm T} ^\ell \cdot E_{\rm T}^{\rm {miss}} \cdot (1-\cos \Delta\phi({\ell,E_{\rm T}^{\rm {miss}}}) ) }$,
is required to be less than 40~\GeV.
$E_{\rm T}^{\rm {miss}}$ is defined 
as the missing transverse momentum based on the calorimeter information
and the transverse momenta of muons within $|\eta|<2.7$~\cite{WZXs}, while
$\Delta\phi({\ell,E_{\rm T}^{\rm {miss}}})$ is the azimuthal angle separation 
between the lepton and the $E_{\rm T}^{\rm {miss}}$ vectors.
Additionally, the following requirements are imposed: 
$\Delta\phi({{\rm jet~or~} e,E_{\rm T}^{\rm {miss}}})\\<0.1$ for the electron control region
and $\Delta\phi({\mu,E_{\rm T}^{\rm {miss}}})<0.5$ for the muon control region. 
For the muon control region, an additional requirement of 
at least one jet is imposed. After these criteria are applied, 
the remaining background from electroweak processes is estimated to be less than 5\%.
The $R_{\rm prompt}$ fractions are measured using $Z$ boson events satisfying $86\GeV <\mll<96\GeV$
via a method referred to as the ``tag-and-probe" method. 
The ``tag'' lepton is required to satisfy all lepton selection criteria, 
while the unbiased opposite charge ``probe'' lepton should satisfy the loose criteria.
The $R_{\rm prompt}$ fractions are parametrised as a function of the lepton  $p_{\rm T}$ and 
range from 89\% to 98\% for muons and 96\% to 99\% for electrons.
To improve the accuracy of the prediction, the fractions are parametrised as a function of kinematic variables
separately for leptons that pass the analysis trigger requirement and those that do not.
The muon $R_{\rm fake}$ is measured separately for muons that originate from heavy flavour jets and those
that do not, where the jet flavour is identified using a combination of the secondary vertex~\cite{SV0}
and impact parameter-based~\cite{JETPROB} $b$-tagging algorithms. 
For muons, $R_{\rm fake}$ ranges from 5\% to 10\% (5\% to 40\%) for heavy flavor (light flavour) jets.
For electrons, $R_{\rm fake}$ ranges from 45\% to 60\%.

A partially data-driven approach is adopted to estimate $Z/\gamma^*\rightarrow ee$ and $Z/\gamma^*\rightarrow \mu\mu$ 
contributions to the OS dilepton channels. A control region is defined requiring $80\GeV <\mll<100\GeV$ and $\ge1$ jets, 
where non-$Z$ boson contributions are found to be negligible. Normalization factors between the observed number of events in data and the 
MC prediction are obtained as a function of jet multiplicity from this region and applied to the MC estimates in the signal region. 
All other backgrounds, including $Z/\gamma^*\rightarrow \tau\tau$, are estimated from MC simulation. 
The contribution of $Z/\gamma^*\rightarrow \tau\tau$ is found to be negligible after all selection criteria are applied.
Table~\ref{Tab:results_OS} summarises the background estimates for the OS channels.
In the OS $ee$ and $\mu\mu$ channels, $Z/\gamma^*$+jets and \ttbar\ backgrounds dominate, while the
\ttbar\ production contributes more than 90\% in the $e\mu$ channel. 
Smaller contributions arise from diboson production and events with fake leptons.

The fraction of reconstructed electrons with charge mis\-identification due to hard bremsstrahlung  
is measured from simulated $Z/\gamma^*$+jets events, by comparing the MC generated charge of the 
electron originating from the $Z$ to that of the reconstructed electron candidate. 
The fraction is parame\-trised as a function of the electron $E_{\rm T}$ and $\eta$ and applied to 
$Z/\gamma^*\rightarrow e^+e^-$ and
$t\bar t \rightarrow e^{\pm}\ell^{\mp} b \bar b$ MC backgrounds to obtain their contributions to the SS dilepton final state, 
thus benefiting from the large number of simulated OS events. 
Since the MC overestimates the charge misidentification as observed in the $Z/\gamma^* \rightarrow ee$ data sample, 
$\eta$-dependent scale factors between data and MC simulation 
are obtained using $Z/\gamma^* \rightarrow e^\pm e^\pm$ events with $80\GeV <\mll<100\GeV$. 
Both electrons are required to be within the same $\eta$ range and with the same charge.
These factors are applied to the MC estimates.
The rate of charge misidentification due to tracking resolution was found to be negligible 
within the lepton transverse momentum range of interest and in good agreement with the MC predictions.

All other backgrounds are estimated from MC simulation and found to be small, as shown in Table~\ref{Tab:results_SS}.  
In the SS $ee$ and $e\mu$ channels, the dominant background arises from fake leptons. The next 
most significant background is diboson production for the $e\mu$ channel and $Z/\gamma^*$ production for the $ee$ channel.
The SS $\mu\mu$ channel is dominated by the diboson background with a smaller contribution from fake leptons.

\section{Systematic uncertainties}
\label{Sec:Sys}

%%%%%%%%%%%%%
% OS/SS RESULTS
%%%%%%%%%%%%%
\begin{table*}[]
\caption{Summary of the expected background yields and observed numbers of events for the OS dilepton channels.
The top part of the table shows the numbers obtained for events with two leptons, $\ge 1$ jet, $\mll>110\GeV$,
and $S_{\rm T}>400~\GeV$.
The bottom part of the table shows the numbers for the final LRSM selection, where 
an additional requirement $\mlljj \ge 400~\GeV$ is imposed.  
The quoted uncertainties include statistical and systematic components,
excluding the luminosity uncertainty of $\pm 3.7\%$. The latter is relevant for all backgrounds except for the 
fake lepton(s) background, which is measured using data. }
\label{Tab:results_OS}

\begin{tabular*}{\textwidth}{@{\extracolsep{\fill}}
l
|r@{\hspace{1pt}}r@{\hspace{1pt}}r
|r@{\hspace{1pt}}r@{\hspace{1pt}}r
|r@{\hspace{1pt}}r@{\hspace{1pt}}r
|r@{\hspace{1pt}}r@{\hspace{1pt}}r@{}
}
  \hline
  \hline
 Physics Processes & \multicolumn{3}{c|}{$e^{\pm}e^{\mp}$} &
  \multicolumn{3}{c|}{$\mu^{\pm}\mu^{\mp}$} &
  \multicolumn{3}{c|}{$e^{\pm}\mu^{\mp}$} &
  \multicolumn{3}{c}{Total} \\ \hline

$Z/\gamma^*$+jets      & 136.1 & $\pm$ & 12.5 & 173.2 & $\pm$ & 15.1 & 0.8 & $\pm$ & 0.8 & 310 & $\pm$ & 20\\
Diboson                          & 4.3 & $\pm$ & 1.8 & 7.3 & $\pm$ & 1.9 & 5.9 & $\pm$ & 1.6 & 18 & $\pm$ & 3\\
Top                      & 103.1 & $\pm$ & 12.3 & 100.9 & $\pm$ & 12.0 & 199.4 & $\pm$ & 23.3 & 403 & $\pm$ & 46\\
Fake lepton(s)                    & 12.5 & $\pm$ & 8.1 & -0.2 & $\pm$ & 0.7 & 6.1 & $\pm$ & 4.2 & 18 & $\pm$ & 9\\
\hline
Total Background       & 256.0 & $\pm$ & 26.2 & 281.2  & $\pm$ & 27.9 & 212.3  & $\pm$ & 33.8 & 750 & $\pm$ & 78\\
Observed events       & 248~~~ &             &          & 245~~~ &             &          & 247~~~ &             &          & 740 &             &\\ \hline
                   & \multicolumn{12}{c}{{\bf $\mlljj \ge 400~\GeV$}} \\ \hline
Total Background & 254.8  & $\pm$ & 25.8     & 279.7  & $\pm$  &  27.6 & 210.9  & $\pm$ & 33.4 & 745 & $\pm$ & 77\\
Observed events  & 246~~~ &             &               & 241~~~ &            &          &  244~~~ &            &          &  731 &            &\\
\hline\hline

\end{tabular*}
\end{table*}

\begin{table*}[]
\caption{Summary of the expected background yields and observed numbers of events for the SS dilepton channels.
The top part of the table shows the numbers obtained for events with two leptons, $\ge 1$ jet and $\mll>110\GeV$.
The bottom part of the table shows the numbers for the final LRSM selection, where 
an additional requirement $\mlljj \ge 400~\GeV$ is imposed.  
The quoted uncertainties include statistical and systematic components,
excluding the luminosity uncertainty of $\pm 3.7\%$. The latter is relevant for all backgrounds except for the 
fake lepton(s) background, which is measured using data. }
\label{Tab:results_SS}
\begin{tabular*}{\textwidth}{@{\extracolsep{\fill}}
l
|r@{\hspace{1pt}}r@{\hspace{1pt}}r
|r@{\hspace{1pt}}r@{\hspace{1pt}}r
|r@{\hspace{1pt}}r@{\hspace{1pt}}r
|r@{\hspace{1pt}}r@{\hspace{1pt}}r@{}
}
  \hline  
  \hline
  Physics Processes & \multicolumn{3}{c|}{$e^{\pm}e^{\pm}$} &
  \multicolumn{3}{c|}{$\mu^{\pm}\mu^{\pm}$} &
  \multicolumn{3}{c|}{$e^{\pm}\mu^{\pm}$} &
  \multicolumn{3}{c}{Total} \\ \hline

$Z/\gamma^*$+jets & 26.1 & $\pm$ & 5.6 & 0.0 &  $^{+}_{-}$  & $^{1.6}_{0}$&  1.2 & $\pm$ & 0.7 & 27 & $\pm$ & 6\\
Diboson & 12.7 & $\pm$ & 2.3 & 7.2 & $\pm$ & 1.7 & 18.8 & $\pm$ & 3.0 & 39 & $\pm$ & 6\\
Top  & 5.8 & $\pm$ & 1.3 & 0.7 & $\pm$ & 0.3 & 6.8 & $\pm$ & 1.6 & 13 & $\pm$ & 3\\
Fake lepton(s) & 93.6 & $\pm$ & 35.7 & 3.1 & $\pm$ & 1.6 & 53.8 & $\pm$ & 20.3 & 151 & $\pm$ & 50\\
\hline
Total Background & 138.3 & $\pm$ & 36.5 & 11.0 & $^{+}_{-}$ & $^{2.9}_{2.5}$ & 80.7 & $\pm$ & 20.8 & 230 & $\pm$ & 52\\
Observed events & 155~~~ & & &    14~~~ &  &  &    99~~~ &  &    &  268&  &   \\ \hline
                   & \multicolumn{12}{c}{{\bf $\mlljj \ge 400~\GeV$}} \\ \hline
Total Background & 48.4 & $\pm$ & 16.1 & 4.4 & $^{+}_{-}$ &  $^{2.1}_{1.3}$ & 24.6 & $\pm$ & 7.6 & 77 & $\pm$ & 21\\
Observed events  & 59~~~ &   &  &  8~~~ &  &  &  39~~~ &  &  & 106&  &  \\
  \hline\hline
\end{tabular*}
\end{table*}

The dominant contribution to the systematic uncertainties in the SS $ee$ and $e\mu$ channels arises from the 
fake lepton background estimate. 
As a first step in validating the parametrisation of $R_{\rm fake}$  and $R_{\rm prompt}$, a 
closure test is performed in data. Measurements of $R_{\rm fake}$  and $R_{\rm prompt}$ are
obtained by randomly sampling half of the control regions. The predicted values are then compared
with the values measured in the other half of the data.
The closure test yields an agreement for $R_{\rm fake}$ and $R_{\rm prompt}$ of  respectively 
$\pm40\%$ and $\pm5\%$ for muons and, for electrons, $\pm5\%$ ($\pm20\%$ for $1<|\eta|<1.9$)
and $\pm2\%$, which are propagated to the fake lepton background estimate.
To evaluate the uncertainty on the overall fake lepton background estimate,
the robustness of the procedure is tested against variations across samples.
The estimated fake lepton background is compared to the observed background in SS
events passing the same selection as events in the signal region but where
the sub-leading lepton has a transverse momentum between 15~\GeV\ and 25~\GeV.
The fake lepton background contributes between 65\% in $e^\pm e^\pm$ to 
87\% in $e^\pm \mu^\pm$ of the total background.
This study tests the reliability of both the parametrisation and the use of
$R_{\rm fake}$ and $R_{\rm prompt}$ to extract the background prediction.
In this sample, the total background prediction agrees with the observed data within $\pm 10\%$. 
A $\pm 30\%$ overall systematic uncertainty is assigned 
to cover for the differences between the predicted and the observed $m_{\ell\ell}$ spectra.

The uncertainties on the background due to the electron charge misidentification arise from the 
limited number of MC events used to parametrise the rate and the scale factors used to correct the simulation 
for differences between data and MC and contribute $\pm13\%$ and $\pm12\%$, respectively.

The background and signal estimates derived from MC are affected by  
the jet energy scale (JES) calibration and the jet energy resolution (JER),
theoretical and MC modelling uncertainties, 
and \pt\ and $\eta$ dependent uncertainties on the lepton identification and reconstruction efficiencies  
(identification $\pm (0.2-3.3)$\%, \pT\ scale $\pm (0.2-2)\%$ and resolution $\pm (0.4-10)\%$) \cite{ElePerf, MuRes, MuId}.
The JES ($\pm (2-6)\%$) and JER ($\pm (5-12)\%$) uncertainties applied depend on jet \pT\ and $\eta$ and 
are measured from the 2010 dataset~\cite{JES}. An additional contribution  of $\pm (2-7)\%$ to the JES 
uncertainty is added in quadrature to account for the effect of high luminosity pile-up in the 2011 dataset. 
MC modelling uncertainties for \ttbar\ production~\cite{TopXs} are derived using different MC generators and 
varying, within their uncertainties, the parameters that control initial and final state radiation. 
The resulting uncertainties are $\pm 15\%$ and $\pm (5-7)\%$ for \ttbar\ and diboson contributions respectively.

Due to the limited knowledge of PDFs and $\alpha_s$, the uncertainties are evaluated using a range of current 
PDF sets with the procedure described in Ref.~\cite{PDFSys}.
The final uncertainty is taken from the outer bounds of the overall error bands.
The PDF uncertainties are estimated to be $\pm 9\%$ for the LRSM signal and $\pm 12\%$ for the HNEO signals.

\FloatBarrier
\section{Results and Interpretation}
\label{Sec:Results}

The expected and observed numbers of events in each dilepton final state for the 
baseline selection and the LRSM selections are compared in Tables~\ref{Tab:results_OS} and~\ref{Tab:results_SS}  
for the OS and SS events, respectively.
Additionally, the reconstructed ma\-sses of the $N$ and \WR\ candidates, 
\mljj\ and \mlljj\ respectively, are examined in each dilepton channel. 
Figures~\ref{fig:Mljj} and \ref{fig:Mlljj} show those distributions for the OS and SS 
channels ($ee$, $\mu\mu$, and $e\mu$ combined). 
  
Given the good agreement between the data and the expectations from SM processes, 
the results are used to set limits at 95\% C.L.  on the
visible cross section, $\sigma \mathcal{A} \epsilon$, 
where $\sigma$ is the cross section for new phenomena, 
$\mathcal{A}$ is the acceptance ({\it i.e.} the fraction of events passing geometric and kinematic selection requirements at the particle level),
and $\epsilon$ is the efficiency ({\it i.e.} the detector reconstruction and identification efficiency).
For the HNEO model, $\mathcal{A} \epsilon$ is about 10\% for $m_N=0.1~\TeV$ and reaches a plateau value of about 
28\% at around $m_N=0.8~\TeV$, for all six dilepton channels.
For the LRSM, $\mathcal{A} \epsilon$ varies between 40\% and 65\% across the ($m_{\WR}, m_N$) plane. 
The lowest $\mathcal{A} \epsilon$ occurs for small $m_N$.
It should be noted that the difference in $\mathcal{A} \epsilon$ between the two models is dominated by the fact the decays to $\tau$ leptons are 
included in generating the HNEO samples, while only decays to $e$ and $\mu$ are included in the LRSM samples.
Table~\ref{Tab:VisXs} quotes the limits obtained for each channel, after the baseline selection.

%%%%%%%%%%%%%
% Mljj
%%%%%%%%%%%%%
\begin{figure}[t!]
\small
 \begin{center}
   \subfigure{\includegraphics[width=.45\textwidth]{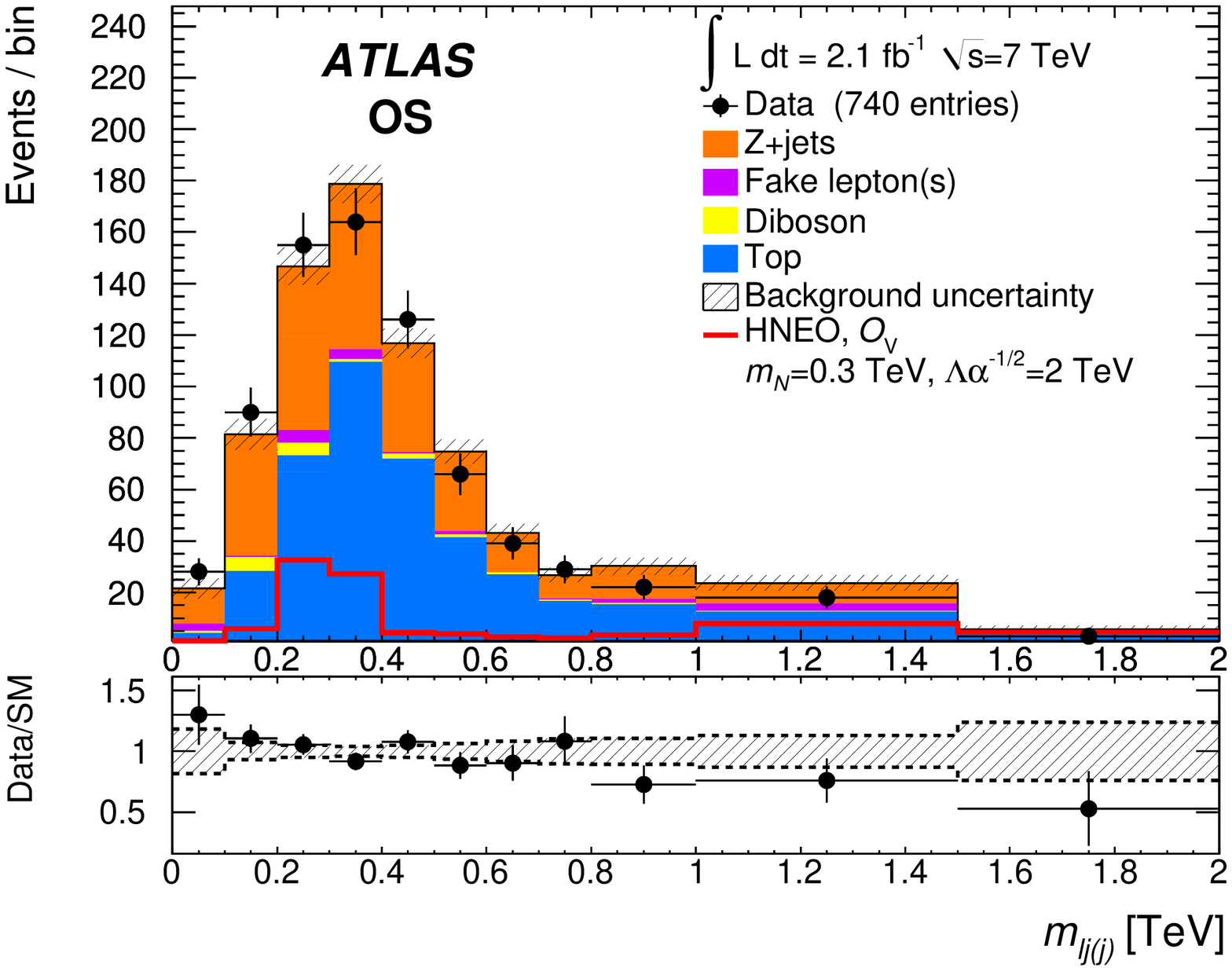}}
   \subfigure{\includegraphics[width=.45\textwidth]{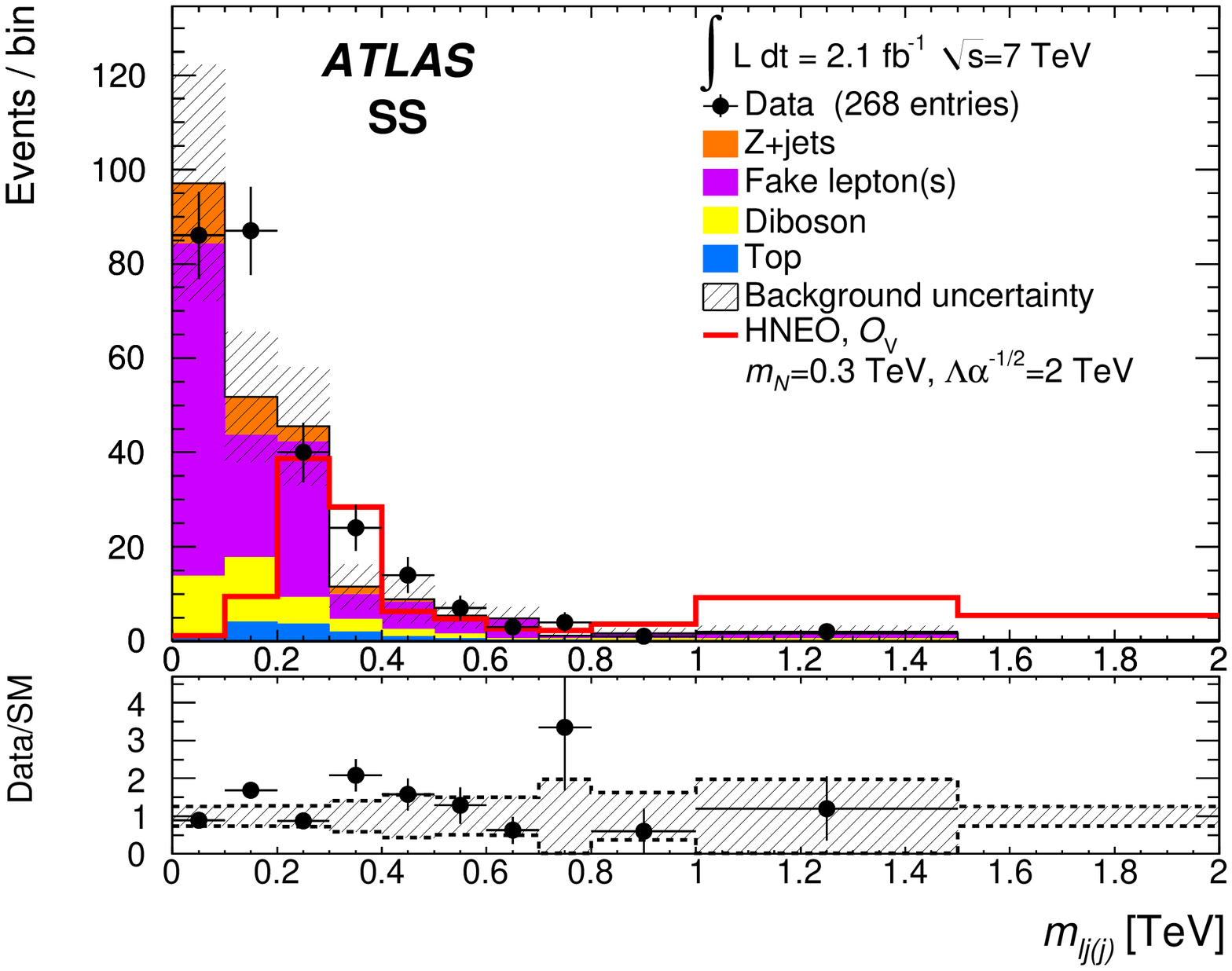}}
 \end{center}
 \caption{Distributions of the reconstructed $N$ invariant mass, \mljj, for  
OS (top) and SS (bottom) dilepton events with $\ge 1$~jets and $\mll>110~\GeV$.
A selection criterion $S_{\rm T}\ge 400~\GeV$ is used for the OS selection.
The hypothetical signal distributions for $m_{N}=0.3\TeV$ for $\mathcal{O}_V$ and $\Lambda/\sqrt{\alpha}$=2~\TeV\ are superimposed.}
 \label{fig:Mljj}
\end{figure}

%%%%%%%%%%%%%
% Mlljj
%%%%%%%%%%%%%
\begin{figure}[t!]
\small
 \begin{center}
   \subfigure{\includegraphics[width=.45\textwidth]{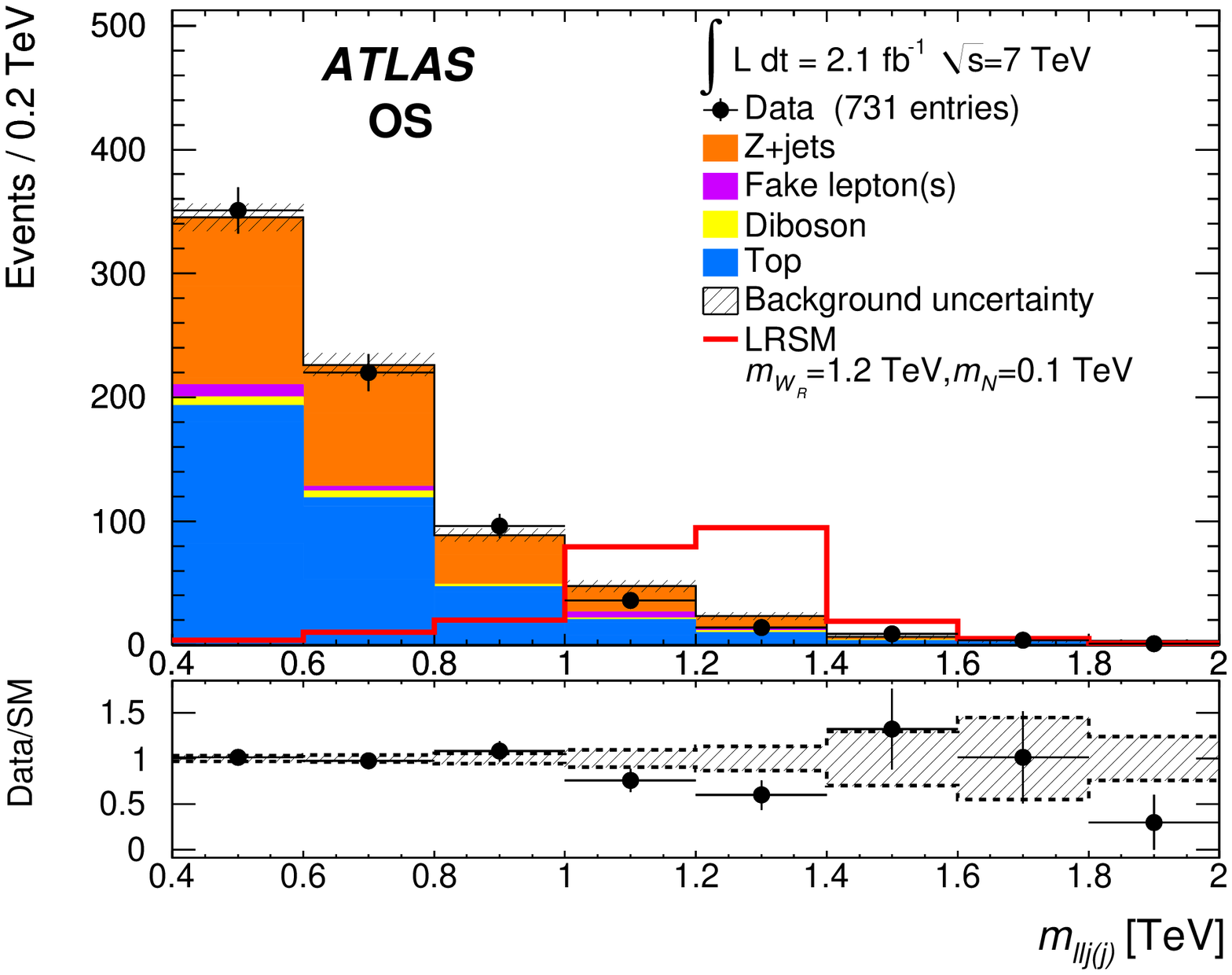}}
   \subfigure{\includegraphics[width=.45\textwidth]{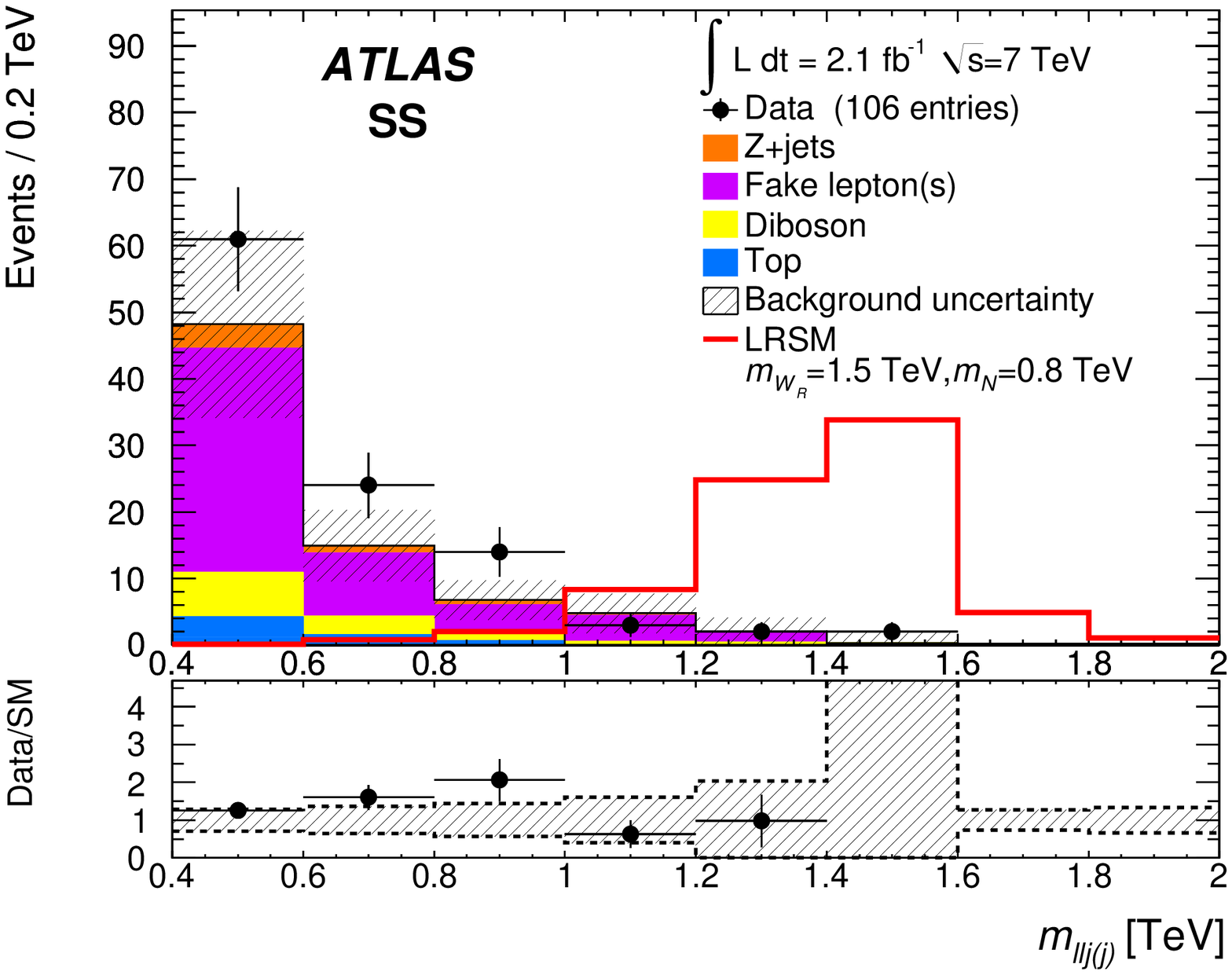}}
 \end{center}
 \caption{Distributions of the reconstructed \WR\ invariant mass, \mlljj, for  
OS (top) and SS (bottom) dilepton events with $\ge 1$~jets, $\mll>110~\GeV$, and $\mlljj \ge 400~\GeV$.
A selection criterion $S_{\rm T}\ge 400~\GeV$ is used for the OS selection.
The hypothetical signal distributions for $m_{\WR}=1.2\TeV$ and $m_N=0.1\TeV$ 
($m_{\WR}=1.5\TeV$ and $m_N=0.8\TeV$)  for the case of maximal mixing, 
are superimposed to the OS (SS) distribution. }
 \label{fig:Mlljj}
\end{figure}

%%%%%%%%%%%%%%%%%%%%%%
% Visible cross section, p-Value (S=0)
%%%%%%%%%%%%%%%%%%%%%%
\begin{table}[b!]
\caption{Observed (${\rm obs}$) and expected (${\rm exp}$) 95\% C.L. upper limits on the visible cross section, $\langle \sigma \mathcal{A} \epsilon \rangle ^{95}$,
for each OS and SS dilepton channel after the baseline selection.
}
\label{Tab:VisXs}                                                                                                             
\begin{tabular*}{\columnwidth}{@{\extracolsep{\fill}}lcc@{}}                                                               
\hline \hline                                                                                                                 
Channels &  
$\langle \sigma \mathcal{A} \epsilon \rangle ^{95}_{\rm obs} $ [fb]                                                                                                          
& $\langle \sigma \mathcal{A} \epsilon \rangle ^{95}_{\rm exp} $ [fb]     \\                                                                                                        
                                                                                                         
\hline                            
$e^{\pm}e^{\mp}$            & 28.6   & 31.0        \\
$\mu^{\pm}\mu^{\mp}$    & 25.1  & 36.7        \\
$e^{\pm}\mu^{\mp}$        & 50.9   & 36.4        \\ \hline
$e^{\pm}e^{\pm}$            & 37.6   & 29.6       \\
$\mu^{\pm}\mu^{\pm}$    &  ~6.1  & 4.6      \\
$e^{\pm}\mu^{\pm}$        & 25.4   & 16.2      \\
\hline\hline                                                                                                                  
\end{tabular*}                                                                                                                
\end{table}

The resulting limits for the interpretation of the data in terms of the HNEO and LRSM models 
are derived using as templates the reconstructed masses of the $N$ and \WR\ candidates in each dilepton channel. 
The baseline selection is used for the HNEO model, while the additional cut of \mlljj\ is applied for the LRSM model.
Systematic uncertainties from JES and JER are included as variations in the signal and background templates.
The uncertainties on the measurement of $R_{\rm fake}$ and $R_{\rm prompt}$ 
are included as variations in the fake lepton background templates.
All other uncertainties have no significant kinematic dependence.
Correlations of uncertainties between signal and background, as well as across channels, are taken into account.

The 95\% C.L. exclusion limits on the mass of the heavy neutrino in the HNEO model 
and their dependence on $\Lambda/\sqrt{\alpha}$ 
are shown in Fig.~\ref{fig:HNEO_Limit} for the Majorana and Dirac scenarios 
using various effective operator hypotheses. 
Figure~\ref{fig:LRSM_Limit} shows the exclusion limits 
for the masses of heavy neutrinos and the \WR\ boson in the LRSM interpretation, 
for the no-mixing and maximal-mixing scenarios between $N_e$ and $N_\mu$ neutrinos, 
for both the Majorana and Dirac heavy neutrinos hypotheses.

The above results are obtained with a Bayesian~\cite{MClimit} approach, 
where systematic uncertainties are treated as nuisance parameters with 
a truncated Gaussian as a prior shape. 
The prior shape on the parameters of interest, $\sigma \times $ BR, is 
assumed to be flat.

%%%%%%%%%%%%%
% Limit HNEO
%%%%%%%%%%%%%
\begin{figure}[t!]
\begin{center}
   \subfigure{\includegraphics[width=.45\textwidth]{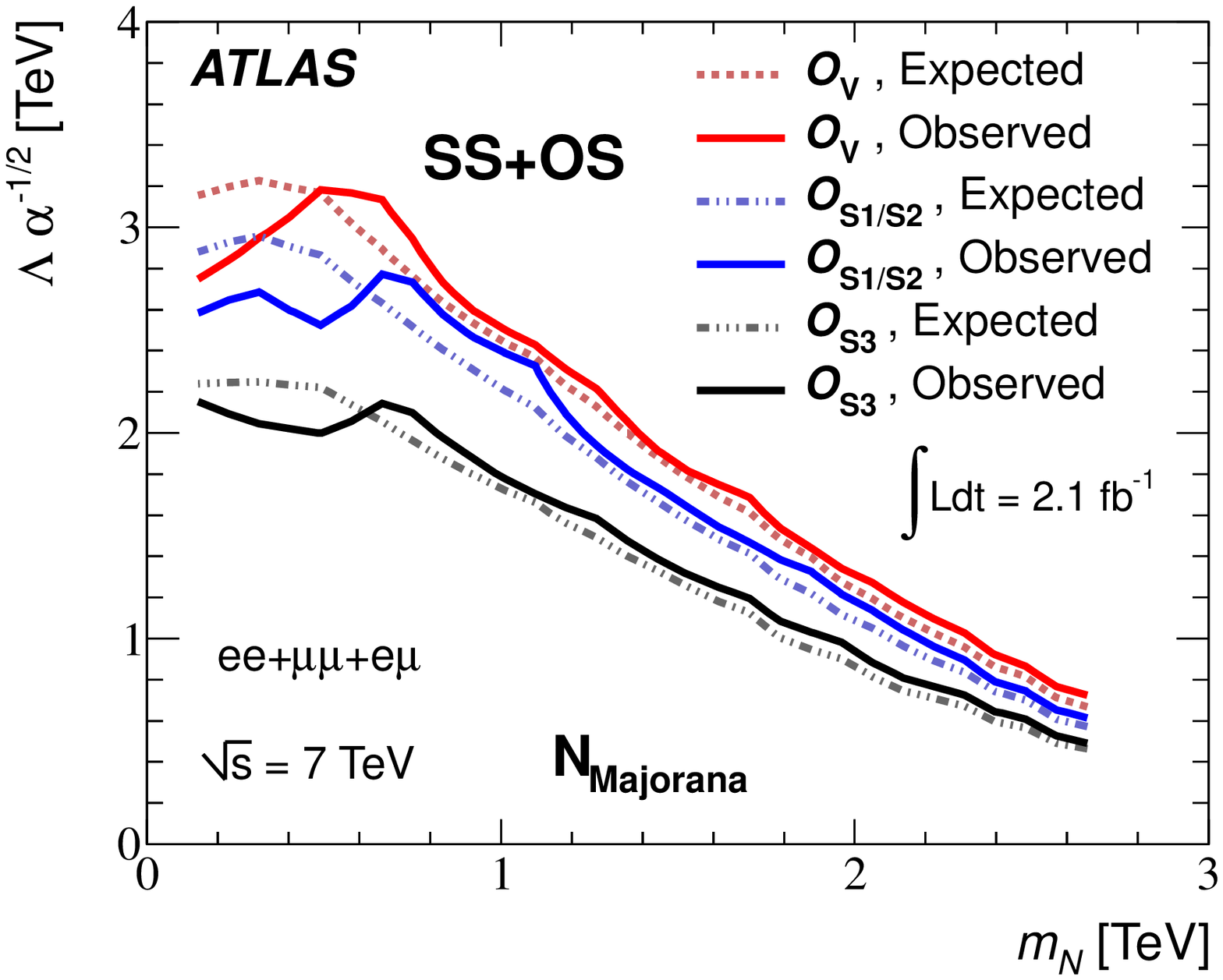}}
   \subfigure{\includegraphics[width=.45\textwidth]{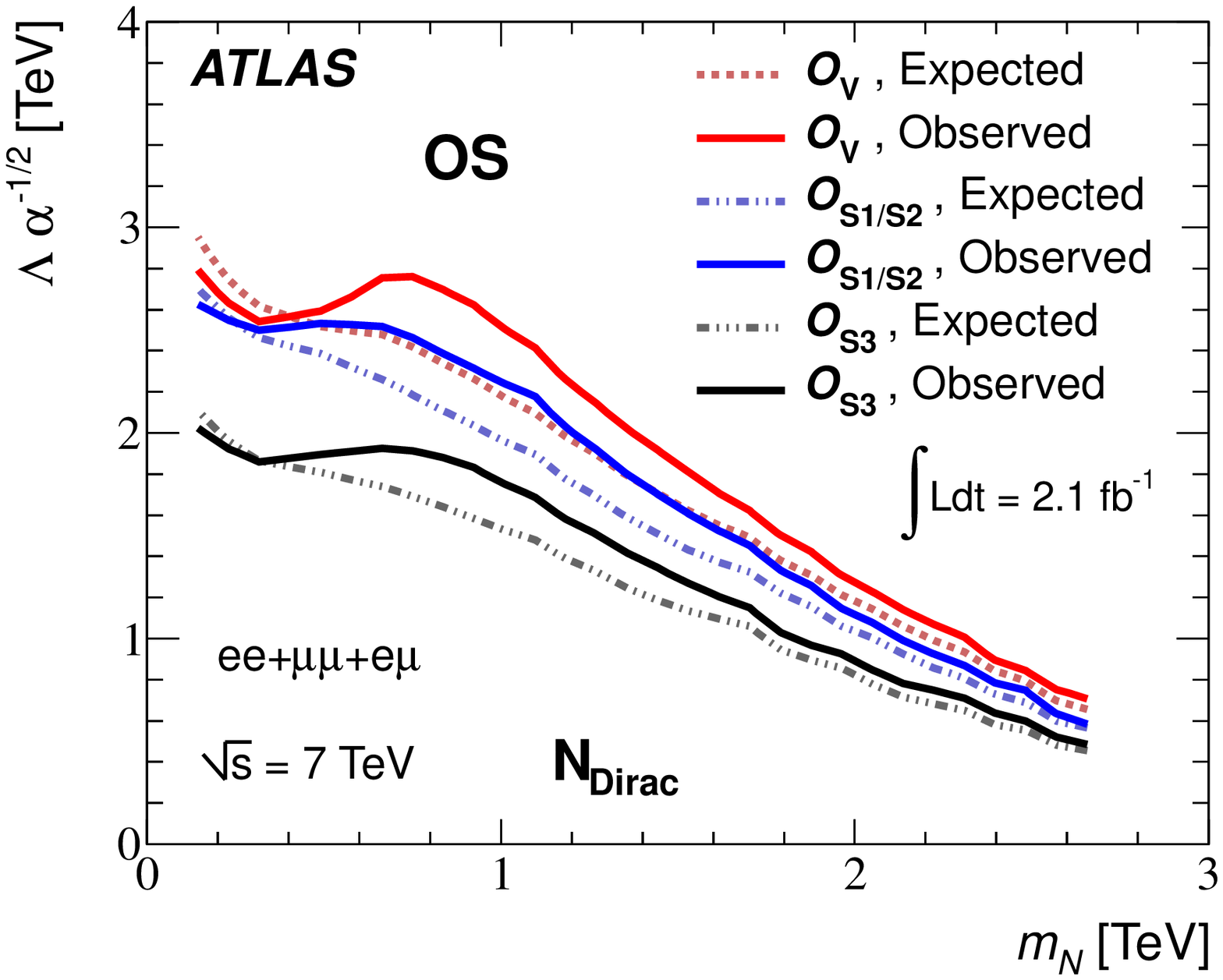}}
 \end{center}
 \caption{Expected and observed 95\% C.L. upper limits on  $\Lambda / \sqrt{\alpha}$ as a function of 
the mass of a heavy neutrino, for the operators $\mathcal{O}_V$, $\mathcal{O}_{s1}$/$\mathcal{O}_{s2}$, and $\mathcal{O}_{s3}$, 
using the formalism of Lagrangian of effective operators, for the Majorana (top) and Dirac (bottom) scenarios.
}
\label{fig:HNEO_Limit}
\end{figure}

%%%%%%%%%%%%%
% Limit LRSM
%%%%%%%%%%%%%
\begin{figure}[t!]
\begin{center}
   \subfigure{\includegraphics[width=.45\textwidth]{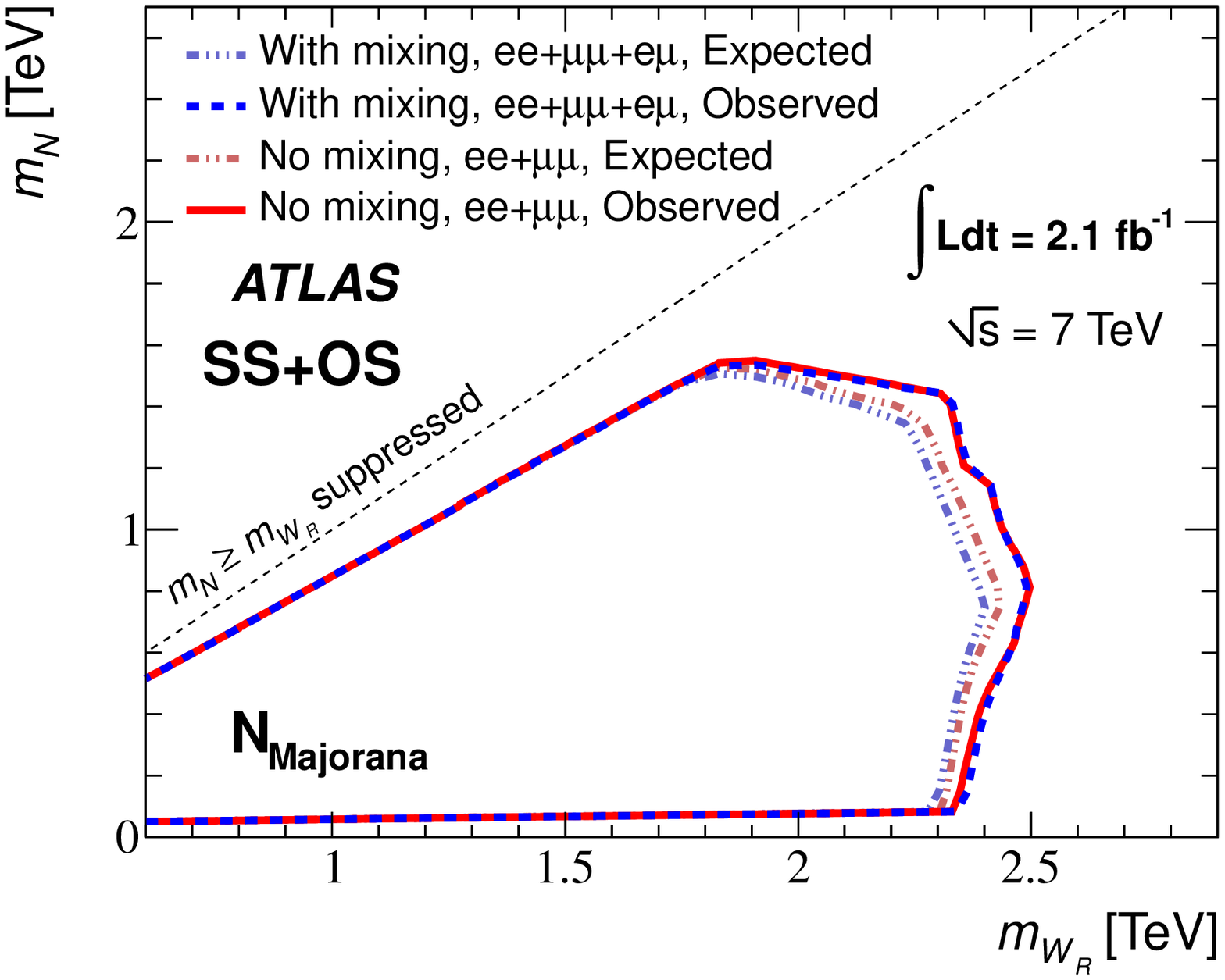}}
   \subfigure{\includegraphics[width=.45\textwidth]{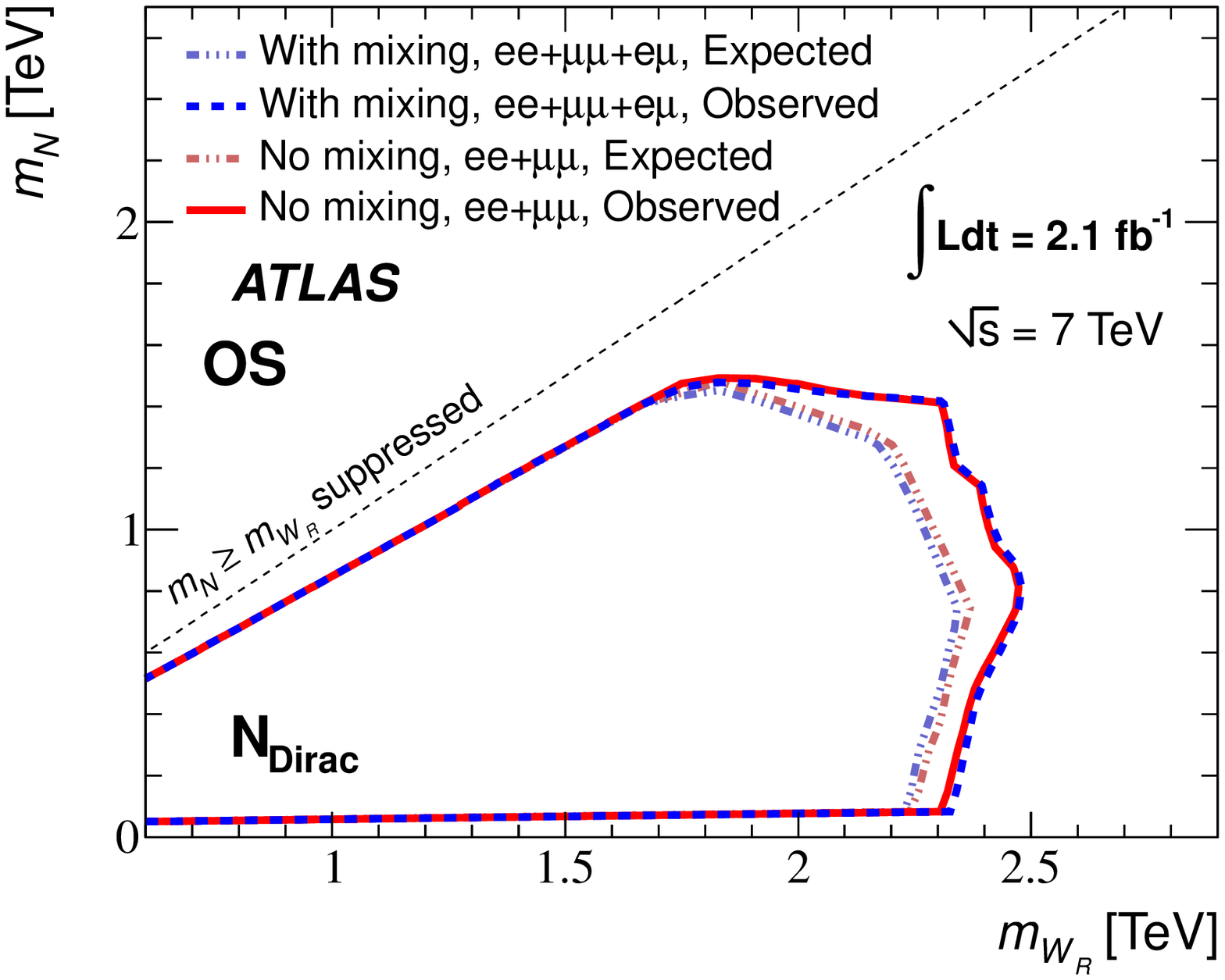}}
 \end{center}
 \caption{Expected and observed 95\% C.L. upper limits on the heavy neutrino and \WR\
masses for the Majorana (top) and Dirac (bottom) cases, in the no-mixing and maximal-mixing scenarios. 
}
\label{fig:LRSM_Limit}
\end{figure}

\section{Conclusions}
\label{Sec:Concl}

A dedicated search for hypothetical heavy Majorana and Dirac neutrinos, 
and \WR\ bosons in final states with two high-\pT\ same-sign or 
opposite-sign leptons and ha\-dronic jets has been presented.
In a data sample corresponding to an integrated $pp$ luminosity of 2.1~\ifb\ 
at $\sqrt{s}$ = 7~TeV, no significant deviations 
from the SM expectations are observed, 
and 95\% C.L. limits are set on the contributions of new physics.
Excluded mass regions for Majorana and Dirac neutrinos 
are presented for various operators of an effective Lagrangian framework
and for the LRSM. 
The latter interpretation was used to extract a 
lower limit on the mass of the gauge boson \WR.
For both no-mixing and maximal-mixing scenarios, \WR\ bosons with masses below 
$\approx1.8~\TeV$ ($\approx2.3~\TeV$) are excluded for mass differences between the \WR\ and $N$ masses larger than 0.3~\TeV\ (0.9~\TeV).  
In the effective Lagrangian interpretation, considering the vector operator and Majorana-type heavy neutrinos,  
the lower limit on $\Lambda / \sqrt{\alpha}$ ranges from $\approx 2.5\TeV$ to $\approx 0.7\TeV$
for heavy neutrino masses ranging from 0.1~\TeV\ to 2.7~\TeV. 
Comparable limits are obtained for Dirac-type neutrinos in both models.
The results described represent the most stringent limits to date 
on the masses of heavy neutrinos and \WR\ boson obtained in direct searches.

% Acknowledgements for papers with collision data
% Version 9-Feb-2012
\section{Acknowledgements}

We thank CERN for the very successful operation of the LHC, as well as the
support staff from our institutions without whom ATLAS could not be
operated efficiently.

We acknowledge the support of ANPCyT, Argentina; YerPhI, Armenia; ARC,
Australia; BMWF, Austria; ANAS, Azerbaijan; SSTC, Belarus; CNPq and FAPESP,
Brazil; \\
NSERC, NRC and CFI, Canada; CERN; CONICYT, Chile; CAS, MOST and NSFC,
China; COLCIENCIAS, Colombia; MSMT CR, MPO CR and VSC CR, Czech Republic;
DNRF, DNSRC and Lundbeck Foundation, Denmark; EPLANET and ERC, European Union;
IN2P3-CNRS, CEA-DSM/IRFU, France; GNAS, Georgia; BMBF, DFG, HGF, MPG and AvH
Foundation, Germany; GSRT, Greece; ISF, MINERVA, GIF, DIP and Benoziyo Center,
Israel; INFN, Italy; MEXT and JSPS, Japan; CNRST, Morocco; FOM and NWO,
Netherlands; RCN, Norway; MNiSW, Poland; \\
GRICES and FCT, Portugal; MERYS
(MECTS), Romania; MES of Russia and ROSATOM, Russian Federation; JINR; MSTD,
Serbia; MSSR, Slovakia; ARRS and MVZT, Slovenia; DST/NRF, South Africa;
MICINN, Spain; SRC and Wallenberg Foundation, Sweden; SER, SNSF and Cantons of
Bern and Geneva, Switzerland; NSC, Taiwan; TAEK, Turkey; STFC, the Royal
Society and Leverhulme Trust, United Kingdom; DOE and NSF, United States of
America.

The crucial computing support from all WLCG partners is acknowledged
gratefully, in particular from CERN and the ATLAS Tier-1 facilities at
TRIUMF (Canada), NDGF (Denmark, Norway, Sweden), CC-IN2P3 (France),\\
KIT/GridKA (Germany), INFN-CNAF (Italy), NL-T1 (Netherlands), PIC (Spain),
ASGC (Taiwan), RAL (UK) and BNL (USA) and in the Tier-2 facilities
worldwide.

%%%%%%%%%%%%%%%%%%%%%%%%%%
% Bibliography
%%%%%%%%%%%%%%%%%%%%%%%%%%
\bibliographystyle{spphys}       
\bibliography{epjc}                     

%%%%%%%%%%%%%%%%%%%%%%%%%%
% ATLAS author list
%%%%%%%%%%%%%%%%%%%%%%%%%%
\clearpage
\onecolumn
% ATLAS Collaboration author list for 19-DEC-2011
% Data extracted on 16-Mar-2012 for paperid 174
%\documentclass[11pt]{article}
%\usepackage{a4wide}\begin{document}
\begin{flushleft}
{\Large The ATLAS Collaboration}

\bigskip

G.~Aad$^{\rm 48}$,
B.~Abbott$^{\rm 110}$,
J.~Abdallah$^{\rm 11}$,
A.A.~Abdelalim$^{\rm 49}$,
A.~Abdesselam$^{\rm 117}$,
O.~Abdinov$^{\rm 10}$,
B.~Abi$^{\rm 111}$,
M.~Abolins$^{\rm 87}$,
O.S.~AbouZeid$^{\rm 157}$,
H.~Abramowicz$^{\rm 152}$
% ...
\bigskip

% ...
$^{10}$ Institute of Physics, Azerbaijan Academy of Sciences, Baku, Azerbaijan\\
$^{11}$ Institut de F\'isica d'Altes Energies and Departament de F\'isica de la Universitat Aut\`onoma  de Barcelona and ICREA, Barcelona, Spain\\
% ...
$^{48}$ Fakult\"{a}t f\"{u}r Mathematik und Physik, Albert-Ludwigs-Universit\"{a}t, Freiburg i.Br., Germany\\
$^{49}$ Section de Physique, Universit\'e de Gen\`eve, Geneva, Switzerland\\
% ...
$^{87}$ Department of Physics and Astronomy, Michigan State University, East Lansing MI, United States of America\\
% ...
$^{110}$ Homer L. Dodge Department of Physics and Astronomy, University of Oklahoma, Norman OK, United States of America\\
$^{111}$ Department of Physics, Oklahoma State University, Stillwater OK, United States of America\\
% ...
$^{117}$ Department of Physics, Oxford University, Oxford, United Kingdom\\
% ...
$^{152}$ Raymond and Beverly Sackler School of Physics and Astronomy, Tel Aviv University, Tel Aviv, Israel\\
% ...
$^{157}$ Department of Physics, University of Toronto, Toronto ON, Canada\\
% ...
\end{flushleft}

%\end{document}

\end{document}